# Astro2020 Science White Paper

# Early evolution of galaxies and of large-scale structure from CMB experiments

**Thematic Areas**: ☐ Planetary Systems  ☐ Star and Planet Formation
☐ Formation and Evolution of Compact Objects  ☒ Cosmology and Fundamental Physics
☐ Stars and Stellar Evolution  ☐ Resolved Stellar Populations and their Environments
☒ Galaxy Evolution  ☐ Multi-Messenger Astronomy and Astrophysics

**Principal Author:**
Name: Gianfranco De Zotti
Institution: INAF-Osservatorio Astronomico di Padova, Italy
Email: gianfranco.dezotti@inaf.it
Phone: +390498293444

**Co-authors:** M. Bonato (INAF-IRA, Bologna, Italy), M. Negrello (Cardiff University, UK), D. Herranz (Universidad de Cantabria, Santander, Spain), M. López-Caniego (ESAC, Villafranca del Castillo, Spain), T. Trombetti and C. Burigana (INAF-IRA, Bologna, Italy), L. Bonavera and J. González-Nuevo (Universidad de Oviedo, Spain), S. Hanany (University of Minnesota), G. Rocha (JPL/Caltech)

**Abstract**: Next generation CMB experiments with arcmin resolution will, *for free*, lay the foundations for a real breakthrough on the study of the early evolution of galaxies and galaxy clusters, thanks to the detection of large samples of strongly gravitationally lensed galaxies and of proto-clusters of dusty galaxies up to high redshifts. This has an enormous legacy value. High resolution follow-up of strongly lensed galaxies will allow the *direct* investigation of their structure and kinematics up to $z \geq 6$, providing direct information on physical processes driving their evolution. Follow-up of proto-clusters will allow an observational validation of the formation history of the most massive dark matter halos up to $z \geq 4$, well beyond the redshift range accessible via X-ray or SZ measurements. These experiments will also allow a giant leap forward in the determination of polarization properties of extragalactic sources, and will provide a complete census of cold dust available for star formation in the local universe.

**Ground-based vs space-borne experiments**

Next generation CMB experiments both from space with ≈ 1.5 m telescopes, like PICO (Hanany et al. 2019), and from the ground, with ≈ 6 m telescopes, like CMB-S4 (Abazajian et al. 2016) and the Simons Observatory (Ade et al 2019) will reach arcmin resolution at sub-mm and mm wavelengths, respectively, with sensitivities at fundamental limits. Space-borne experiments will cover a very broad frequency range, from ≈20 to ≈800 GHz, while observations from the ground are only possible, in atmospheric windows, up to ≈300 GHz. The potential of ground based experiments for extragalactic astrophysics has been demonstrated by SPT (Mocanu et al. 2013) and ACT (Marsden et al. 2014) surveys; next generation experiments will greatly extend the sky coverage. Space-borne experiments with telescopes of size similar to *Planck*'s can reach much fainter flux densities: for state-of-the-art instruments the detection limit is not set by instrumental noise but by confusion noise which scales roughly as the beam solid angle (see Fig. 3 of De Zotti et al. 2015). Thus improving the resolution from ≈5' (*Planck*) to the diffraction limit for a ≈ 1.5 m telescope (≈1' at ≈800 GHz) decreases the detection limit by about one order of magnitude.

**Early phases of galaxy evolution**

CMB experiments with arcmin resolution both from the ground and from space will drive a real breakthrough in the study of early evolutionary phases of galaxies, paving the way to answer major, still open problems like: which are the physical mechanisms shaping the galaxy properties (Silk & Mamon 2012; Somerville & Davé 2015): in situ processes? interactions? mergers? cold flows? How feedback processes work? To settle these issues we need direct information on the structure and the dynamics of high-z galaxies. But these are compact, with typical sizes of 1−2 kpc (e.g., Fujimoto et al. 2018), corresponding to angular sizes of 0.1−0.2 arcsec at z ≈ 2−3. Thus they are hardly resolved even by ALMA and by the HST. If they are resolved, high enough S/N ratios per resolution element are achieved only for the brightest galaxies, not representative of the general population.

Strong gravitational lensing provides a solution. CMB surveys *will detect the brightest sub-mm strongly lensed galaxies in the sky*, with extreme magnifications, $\mu$, up to several tens (Cañameras et al. 2015). Since lensing conserves surface brightness, the effective angular size is stretched by an average factor $\mu^{1/2}$, substantially increasing the resolving power. A spectacular example are ALMA observations of PLCK_G244.8 +54.9 at z≈3.0 with $\mu$≈30 (Cañameras et al. 2017a) which reached the astounding spatial resolution of ≈60 pc, substantially smaller than the size of Galactic giant molecular clouds. Cañameras et al. (2017a) also obtained CO spectroscopy with an uncertainty of 40—50 km/s. This spectral resolution makes possible a direct investigation of massive outflows driven by AGN feedback at high z. In this way Spilker et al. (2018) were able to detect, in a strongly lensed galaxy at z=5.3, a fast (800 km/s) molecular outflow carrying mass at a rate close to the SFR, thus removing a large fraction of the gas available for star-formation. Information on the the energetics of the feedback can come from measurements of the thermal Sunyaev-Zeldovich effect induced by it (Lacy et al. 2019).

The high redshifts of magnified galaxies imply high redshifts of foreground lenses. Optical follow-up will allow us to investigate the total (visible and dark) mass of the lensing galaxies, their density profiles, dark matter sub-structures at higher redshifts than in the case of optical selection (Cañameras et al. 2017b). Also next generation CMB experiments with arcmin resolution will explore essentially the entire Hubble volume for the most intense hyperluminous starbursts, testing whether there are physical limits to the star-formation rates of galaxies.

*Herschel* has demonstrated that at 500 μm (600 GHz) strongly lensed galaxies with $S_{500\mu m} \geq 100$ mJy, close to PICO's detection limit, amount to ≈25% of the total counts; a similar fraction was found at the completeness limit of the SPT survey at 220 GHz (Fig. 1). Such high fractions are an exclusive property of (sub-)mm surveys: searches in other wavebands have yielded fractions ≈0.1% (York et al. 2005, Jackson 2008, Treu 2010). The other main populations contributing to the counts are radio sources, easily identified thanks to their very different spectra or by cross-match with low-frequency all-sky catalogs, and local dusty galaxies, also easily recognized.

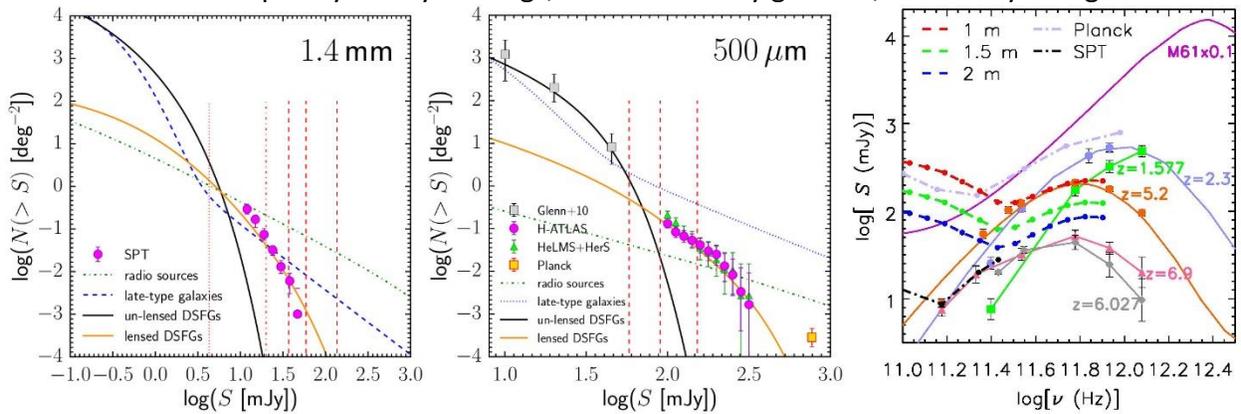

Figure 1. Integral counts of extragalactic sources at 1.4 mm and at 500 μm yielded by the Cai et al. (2013) model, compared with: 1.4 mm SPT counts of candidate strongly lensed galaxies (Mocanu et al. 2013); total 500μm faint Herschel counts (Glenn et al. 2010) and counts of candidate strongly lensed galaxies (Nayyeri et al. 2016, Negrello et al. 2017a); our estimate of Planck counts of strongly lensed galaxies at 545 GHz. The vertical dashed lines show, from right to left, the estimated 5σ confusion limits for diffraction limited surveys with 1, 1.5 and 2 m telescopes. The dot-dashed and dotted lines on the left panel show, respectively, the SPT completeness (Mocanu et al. 2013) and 5σ confusion limits. The right panel shows the observed SEDs of the strongly lensed galaxies SDP 9 (z=1.577; Negrello et al. 2010), SMMJ2135-0102 (z = 2.3259; Swinbank et al. 2010), HLS J091828.6+514223 (z=5.2; Combes et al. 2012), HATLAS J090045.4 +004125 (z=6.027; Zavala et al. 2018) and SPT-S J031132−5823.4 (z=6.9; Strandet e al. 2017). The SED of the local star-forming galaxy M61, scaled down by a factor of 10, is also shown for comparison. The dashed lines on the right panel show the estimated detection limits of space-borne experiments with 1 m , 1.5 m and 2 m telescopes. The dot-dashed black and magenta lines show, respectively, the 90% completeness limits of the Second Planck Catalogue of Compact Sources (Planck Collaboration XXVI 2016) and the SPT completeness limits (Mocanu et al. 2013) extrapolated to 40 and 270 GHz.

The strong difference in redshift between local and strongly-lensed galaxies implies clearly different sub-mm colors, making the separation easy, directly on survey data in the case of space experiments. This is also specific to searches in this frequency range. Selections in other wavebands need spectroscopy or other ancillary data. This is true also for ground-based surveys which provide photometry in the Rayleigh-Jeans region where colors are largely redshift-independent. The right-hand panel of Fig. 1 shows that both ground-based and space-borne experiments like PICO will detect the strongly lensed galaxy HLS J091828.6+514223 at z=5.2. Galaxies like it would be detectable up to z of at least 10. Ground-based and space-borne experiments cover complementary redshift ranges. The former are more efficient at z ≥ 2, the latter at lower z. Space-borne experiments will measure a large fraction of the SEDs, reaching close to the dust emission peak especially at high z, thus allowing us not only to pick up strongly lensed galaxies with close to 100% efficiency (Negrello et al. 2010) but also to get redshift estimates (Gonzalez-Nuevo et al. 2012, Pearson et al. 2013) and to measure bolometric luminosities and dust properties (emissivity index, optical depth, mass, temperature). Yet another important peculiarity of mm/sub-mm surveys is that while the lensed galaxies are heavily dust enshrouded, hence bright at sub-mm wavelengths but faint in the optical, the foreground lenses are mostly massive early-type galaxies in passive evolution, hence optically bright but almost invisible in the sub-mm. This means that there is no, or only weak mutual contamination of the images.

How many strongly lensed galaxies will be detected? *Herschel* extragalactic surveys have yielded a surface density of ≈0.16 deg$^{-2}$ for $S_{500\mu m}$ ≥ 100 mJy (Negrello et al. 2017a). A ≈ 1.5 m telescope will reach a slightly fainter flux density limit over the full sky (excluding the region around the Galactic plane), thus achieving the detection of several thousands strongly lensed galaxies. The number of detections decreases rapidly for smaller telescope sizes; a telescope substantially smaller than 1 m cannot do any better than *Planck*. A 2 m telescope would increase the number of strongly lensed detections by a factor ≈2.5, but their selection would require follow-up to distinguish them from unlensed high-z galaxies dominating the counts at the corresponding flux density limit. An SPT-like ground-based experiment would have a confusion limit of 4-5 mJy at 1.4 mm (left panel of Fig. 1). The surface density of strongly lensed galaxies at this limit is similar to that achieved by a 2 m telescope at 500 μm; at the SPT completeness limit (≈ 20 mJy at 220 GHz/1.4 mm; Mocanu et al. 2013) it is of ≈0.05 deg$^{-2}$ .
Samples of thousands of strongly lensed galaxies offer exciting prospects, both on the astrophysical and on the cosmological side (Treu 2010; Eales 2015):
- They allow us to determine the spatial distribution of dark plus visible mass in galaxies acting as deflectors down to kpc and sub-kpc scales where the poorly understood transition from baryon to dark matter domination occurs. With a sufficient statistics we may address questions like: how do luminous and dark matter density profiles depend on galaxy mass, type, and cosmic time? Are dark matter density profiles and the abundance of dark matter sub-halos consistent with predictions in the framework of the cold dark matter scenario?
- They will allow us to detect faint active nuclei at the centers of lensed galaxies, thus helping to understand the connection between accretion rate and star-formation rate.
- The lens statistics provides information on cosmological parameters, such as the normalization of the spectrum of primordial perturbations, $\sigma_8$, $\Omega_{matter}$ and $\Omega_\Lambda$, and on the equation of state of dark energy.

This has triggered intensive searches in several wavebands (e.g., Treu 2010; Jacobs et al. 2019). The (sub-)mm selection not only provides very easily large samples, but is unique in allowing the investigation of early phases of galaxy evolution, when spheroidal galaxies were forming most of their stars, before becoming bright, or even simply detectable, in the optical.

**Early phases of cluster evolution**
Next generation CMB experiments with arcmin resolution will open a new window on the study of early phases of cluster formation, when their member galaxies were actively star forming and before the hot intergalactic medium was in place. Traditional approaches to cluster detection (X-ray and Sunyaev-Zeldovich (SZ) surveys, galaxy red sequences) have yielded only a handful of confirmed proto-cluster detections at z≥ 1.5 (Overzier 2016). The reason is that, while at z ≤1−1.5 the cluster cores are dominated by passive early-type galaxies and are filled by hot gas, at higher z cluster members enter the dust-obscured star-formation phase and the intergalactic gas is no longer necessarily at the virial temperature. Four spectroscopically confirmed sub-mm proto-clusters at z from 2.4 to 4.3 have been reported (Ivison et al. 2013; Wang et al. 2016; Oteo et al. 2018; Miller et al. 2018; right panel of Fig. 2). *Planck* has demonstrated the power of CMB surveys for the study of large-scale structure (Planck Collaboration XXXIX 2016) but its resolution was too poor to detect individual proto-clusters (Negrello et al. 2017b) whose typical sizes are of ≈1' (cf. also Alberts et al. 2014).

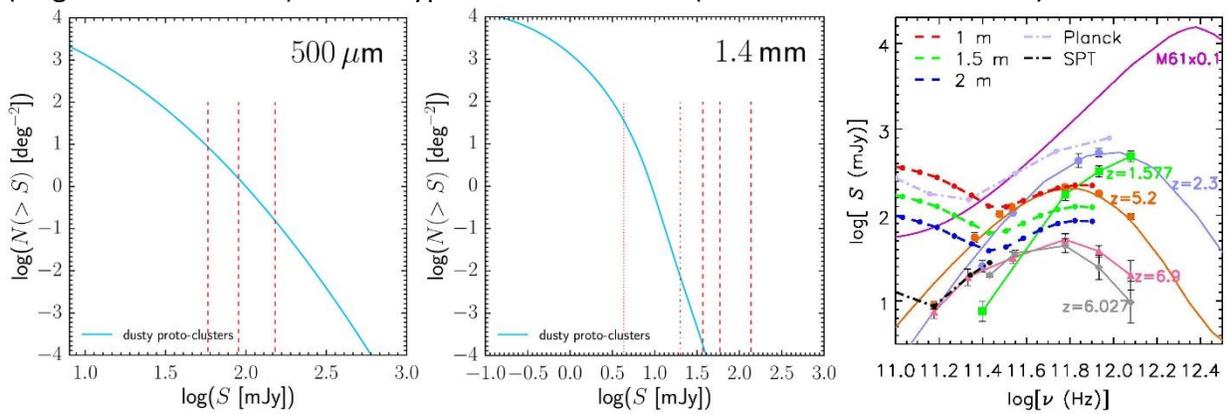

Figure 2. **Left and central panels**. Integral counts of proto-clusters at 500 μm and at 1.4 mm predicted by Negrello et al. (2017b). The dashed vertical red lines in the left and central panels show, from left to right, the 5σ confusion limits for 2, 1.5 and 1 m telescopes operating at the diffraction limit. The dot-dashed and dotted red lines in the central panel correspond to the SPT completeness limit and 5σ confusion limit, respectively. **Right panel**: SEDs of spectroscopically confirmed sub-mm bright high-z proto-clusters compared with the confusion limits of space-borne telescopes and with the SPT completeness limit. The flux densities by Ivison et al. (2013) and Wang et al. (2016) are shown as lower limits because they include the contributions of member galaxies across a ≈100 kpc region, substantially smaller than the expected proto-cluster size (≈ 500 kpc).

Space-borne experiments like PICO will detect many tens of thousands of these objects, at z at least out to ≈4.3 (Fig. 2). This will allow the observational validation of the formation history of the most massive dark matter halos, a crucial test of models for structure formation. Follow-up observations will probe the galaxy evolution in dense environments. Ground based instruments will preferentially detect propto-clusters at the highest redshifts. Since these are expected to be rare, the number of detections will be of only ≈$10^{-2}$ deg$^{-2}$ (central panel of Fig. 2). Detections rapidly increase with decreasing detection limit, reaching ≈40 deg$^{-2}$ at the SPT confusion limit.

**Local dusty galaxies**

We know from *Herschel* that for $S_{500\mu m}$ ≥ 100 mJy the surface density of local dusty galaxies is ≈ 1 deg$^{-2}$ and that their redshift distribution peaks at z=0.02− 0.03 (Negrello et al. 2017a}. A survey with a space borne ≈1.5 m telescope will detect tens of thousands of them. By the time when next generation CMB experiments will fly several wide-angle redshift and photometric surveys will be available, providing distance information for the majority, if not all of them (De Zotti et al. 2018). Combining these data with available or forthcoming data in different wavebands (radio, IRAS, AKARI, WISE, Euclid, GALEX, ROSAT, eROSITA ...) it will be possible to determine, for each galaxy type and as a function of stellar mass, the distribution of dust temperatures and masses, the SFR function, the relationship between star formation and nuclear activity, the contributions of newborn and evolved stars to dust heating, and more. The sample of local galaxies will also be large enough for clustering studies, i.e. to relate the properties of galaxies to the underlying dark matter field and to the properties of their dark matter haloes, as well as to investigate the link between galaxies of different types and their environments. The surface density of local galaxies at the SPT completeness limit is of $\approx 0.05$ deg$^{-2}$ at 220 GHz (Fig. 1), i.e. a factor of $\approx 20$ lower than achieved by a space-borne 1.5 m telescope at 600 GHz. However, as shown by the SED of M61 scaled down by a factor of 10, ground based surveys will detect, for a substantial fraction of low-z galaxies, also the radio emission powered by star formation which is a measure of the SFR unaffected by dust absorption. At a few mm wavelengths a large fraction of the radio emission is free-free, which scales directly with the ionizing luminosity from young stars.

**Polarization**

Next generation CMB experiments will make a giant leap forward in the determination of polarization properties of extragalactic sources. This is because the confusion limit is much lower than in total intensity since it roughly scales as the mean polarization fraction which is of a few percent. Hence the detection limit in polarization is mostly determined by the sensitivity of the instruments, except for telescope sizes substantialy smaller than 1 m. This means that space-borne instruments not only provide measurements over a much broader frquency range but are competitive or better than the ground-based ones also in the common frequency range.

The (sub-)mm polarization properties of dusty galaxies are highly uncertain. The only published measurement (Greaves & Holland 2002) yielded a polarization fraction $\Pi \approx 0.4\%$ at 850 μm for the prototype starburst galaxy M82. From the *Planck* dust polarization maps of the Milky Way, De Zotti et al. (2018) estimated $\Pi \approx 1.4\%$, averaging over inclination. By applying the stacking techniques to a large sample of dusty galaxies drawn from the PCCS2 857 GHz catalogue, Bonavera et al. (2017) estimated a median $\Pi$ of $(2.0 \pm 0.8)\%$ at 353 GHz consistent with the 90% confidence upper limit of 2.2% derived by Trombetti et al. (2018).

The simulations of De Zotti et al. (2018) showed that an experiment with a 1.5 m telescope with the sensitivity of the CORE project will detect, at 800 GHz and |b|>20°, the polarized emission of about 15,000 dusty galaxies at 5σ, down to a flux density of $\approx 6$ mJy, under the conservative assumption of a log-normal probability distribution of the polarization fraction with mean of 0.5% and dispersion of 1%. The number of expected detections decreases rapidly with decreasing frequency. For the assumed probability distribution only $\approx 1$ dusty galaxy per sr is expected at 220 GHz. If so, only space-borne experiments can accurately characterize the contamination of CMB polarization maps by dusty galaxies. Measurements of their polarization properties are informative on the structure and on the ordering of their large-scale magnetic fields.